\documentclass[aps,reprint,amsmath,amssymb,groupaddress,superscriptaddress,pra]{revtex4-2}

\usepackage[utf8]{inputenc}
\usepackage{tcolorbox}
\usepackage{amsmath}
\usepackage{amsfonts}
\usepackage{amssymb}
\usepackage{bm}
\usepackage{float}
\usepackage{xcolor}
\usepackage{graphicx}
\usepackage{soul}
\usepackage{textcomp}
\usepackage{physics}
\usepackage{multirow}
\usepackage{xcolor}
\usepackage{mathtools}
\usepackage{lipsum}
\usepackage{comment}
\usepackage{booktabs}
\usepackage{textcomp}

\usepackage[colorlinks,citecolor=blue,linkcolor=red,anchorcolor=blue,urlcolor=blue]{hyperref}

\begin{document}
\title{Finite-time and Finite-size scalings of coercivity in dynamic hysteresis}
        
\author{Miao Chen}
\affiliation{School of Physics and Astronomy, Beijing Normal University, Beijing, 100875, China}
\affiliation{Key Laboratory of Multiscale Spin Physics (Ministry of Education), Beijing Normal University,Beijing 100875, China}

\author{Xiu-Hua Zhao}
\email{xhzhao@mail.bnu.edu.cn}
\affiliation{School of Physics and Astronomy, Beijing Normal University, Beijing, 100875, China}
\affiliation{Key Laboratory of Multiscale Spin Physics (Ministry of Education), Beijing Normal University,Beijing 100875, China}

\author{Yu-Han Ma}
\email{yhma@bnu.edu.cn}
\affiliation{School of Physics and Astronomy, Beijing Normal University, Beijing, 100875, China}
\affiliation{Key Laboratory of Multiscale Spin Physics (Ministry of Education), Beijing Normal University,Beijing 100875, China}
\affiliation{Graduate School of China Academy of Engineering Physics, Beijing, 100193, China}

\begin{abstract} The coercivity panorama for characterizing the dynamic hysteresis in interacting systems across multiple timescales is proposed by Chen \textit{et al.} in a companion paper. For the stochastic $\phi^4$ model under periodic driving of rate $v_H$, the coercivity landscape $H_c(v_H)$ exhibits plateau features at a characteristic rate $v_P$ with the corresponding coercivity $H_P$. Below this plateau ($v_H<v_P$), the $H_c\sim v_H$ scaling obtained in the near-equilibrium regime becomes inaccessible in the thermodynamic limit. Above the plateau  ($v_H>v_P$), scaling in the fast-driving regime, $H_c\sim v_H^{1/2}$, is completely different from that, $H_c-H_P\sim (v_H-v_P)^{2/3}$, in the post-plateau slow-driving regime. The emergence of the plateau with a finite-size scaling reflects the competition between the thermodynamic limit and the quasi-static limit. In this paper, we provide detailed analytical proofs and numerical evidence supporting these results. Moreover, to demonstrate the coercivity panorama in concrete physical systems, we study the magnetic hysteresis in the Curie-Weiss model and analyze its finite-size effects. We reveal that finite-time coercivity scaling shows model-specific behavior only in the fast-driving regime, while exhibiting universal characteristics elsewhere.

\end{abstract}

\maketitle

\section{Introduction}

Hysteresis refers to the lag in a system's response relative to its equilibrium state under an external driving field. It is characterized by irreversibility and history dependence~\cite{chakrabartiDynamicTransitionsHysteresis1999a,morrisWhatHysteresis2012}. Generally, one anticipates the system to follow the equilibrium curve in the quasistatic limit, where the external field changes infinitely slowly. However, this expectation fails in systems undergoing a first order phase transition (FOPT)~\cite{jungScalingLawDynamical1990,brodyInformationGeometryVapour2008}. In the thermodynamic limit, there are no fluctuations to induce transitions between metastable and stable phases, causing the system to become trapped in a metastable state exhibiting local equilibrium~\cite{moriAsymptoticFormsScaling2010}.

Numerous studies have explored the scaling behavior of hysteresis with respect to the driving rate of the external field, both when quasistatic hysteresis is present and absent. In most cases, the area enclosed by the hysteresis loop serves as a quantitative measure of hysteresis~\cite{jungScalingLawDynamical1990,raoMagneticHysteresisTwo1990,jiangScalingDynamicsLowfrequency1995,sidesKineticIsingModel1999,mooreMesofrequencyDynamicHysteresis2004,gengUniversalScalingDynamic2020}. In systems without quasistatic hysteresis, a linear scaling of the hysteresis area is a common result in the near-equilibrium regime~\cite{goldszteinDynamicalHysteresisStatic1997}. In contrast, for systems undergoing FOPT with quasistatic hysteresis, it has been shown that the difference in hysteresis area between the finite-time case and quasi-static limit scales as $2/3$-power-law scaling in the slow-driving regime~\cite{jungScalingLawDynamical1990,luseDiscontinuousScalingHysteresis1994}. Interestingly, both types of scaling have been observed within the same model under different conditions. Wu \textit{et al.}~\cite{wuErgodicityBreakingScaling2025} studied the scaling relation of extra work in the Curie-Weiss model and revealed scaling crossover that reflects the interplay between finite-time dynamics and finite-size effects. Moreover, numerical and experimental investigations found a richer diversity of scaling behaviors of dynamic hysteresis~\cite{mooreMesofrequencyDynamicHysteresis2004,banerjeeFinitedimensionalSignatureSpinodal2023,kunduDynamicHysteresisNoisy2023}.

For systems exhibiting FOPT but away from the thermodynamic limit and subject to finite fluctuations, fully characterizing the global landscape of hysteresis scaling across different timescales remains challenging. This difficulty arises from the complex interplay among multiple characteristic timescales and the inherent ambiguity in distinguishing between local and global equilibrium regimes~\cite{mahatoHysteresisRateCompetition1992,rikvoldMetastableLifetimesKinetic1994,moriAsymptoticFormsScaling2010,wuErgodicityBreakingScaling2025}.
Moreover, while dynamical phase transitions (DPT) inspired by fast-driving hysteresis phenomena have attracted significant attention~\cite{tomeDynamicPhaseTransition1990,chakrabartiDynamicTransitionsHysteresis1999a,robbEvidenceDynamicPhase2008,riegoUnderstandingDynamicPhase2018}, the transition regime from slow-driving FOPT to DPT has been largely overlooked, and whether a universal scaling exists in this intermediate regime remains an open question. We bridge these gaps in a companion paper~\cite{CP} by proposing the concept of coercivity panorama. We have fully displayed the finite-time and finite-size scaling behaviors of hysteresis across multiple timescales and system sizes in the stochastic $\phi^4$ model~\cite{raoMagneticHysteresisTwo1990,kunduDynamicHysteresisNoisy2023}. In this paper, we not only provide extensive analytical and numerical support for these results in a self-contained way, but also study the coercivity landscape in the Curie-Weiss model, clarifying universal scalings and model-specific scalings in the coercivity panorama.

The rest of this paper is organized as follows. In Sec.~\ref{sec:stochastic_phi4}, we introduce the stochastic $\phi^4$ model, specify its Langevin dynamics under a periodic driving field, and extract its hysteresis loops and coercivity from numerical simulations. In Sec.~\ref{sec:coercivity_stochastic_phi4}, we derive the exact evolution equation for the ensemble‐averaged order parameter, from which the finite-time scaling relations of coercivity in different timescales  are obtained. Moreover, we investigate the finite-size scaling of the coercivity plateau by utilizing renormalization-group theory. To demonstrate the advantages of the coercivity panorama in studying hysteresis phenomena in concrete microscopic models, in Sec.~\ref{sec:coercivity_panorama_CW}, we investigate the magnetic hysteresis of the Curie–Weiss model. A universal collapse of coercivity across system sizes and model‑specific finite-time scalings in the fast-driving regime are demonstrated. 
Conclusions are drawn in Sec.~\ref{sec:conclusion}.

\section{Stochastic $\phi^4$ model}
\label{sec:stochastic_phi4}  

The stochastic $\phi^4$ model describes the dynamics of a system with a double-well potential subjected to external driving and thermal fluctuations. In the thermodynamic limit, the deterministic $\phi^4$ model exhibits a sharp first-order phase transition (FOPT)  where the system's order parameter changes discontinuously as the external field crosses a critical value. To capture the finite-size effects, a noise term is introduced into the system's dynamics to represent finite fluctuations and the resulting stochastic behavior away from the thermodynamic limit~\cite{tuckerOnsetSuperconductivityOneDimensional1971,raoMagneticHysteresisTwo1990}. By incorporating stochastic dynamics, the model allows for the exploration of dynamic hysteresis in finite systems while preserving the key features of first-order transitions, making it an ideal platform for understanding the interplay between finite-time driving and finite-size effects.

In this section, we investigate the dynamics of the stochastic $\phi^4$ model under the driving of an external field and further demonstrate its hysteresis loops under periodic driving.
\subsection{Dynamics}
For a system of interest described by the Landau model interacting with an external field $H$, its free energy density, as a function of the dimensionless order parameter $\phi$, follows~\cite{reichlModernCourseStatistical2016}
\begin{equation}
\label{eq:phi4-free-energy}
    f_4(\phi, H) = \frac{1}{2} a_2 \phi ^2 + \frac{1}{4} a_4 \phi^4 - H \phi,
\end{equation}
Here the constants $a_2$ and $a_4$ correspond to the reduced temperature and intrinsic interaction strength, respectively. We focus on the case $a_2<0$, for which the system exhibits spontaneous symmetry breaking in the absence of both external field and fluctuations. 
The free energy landscape exhibits $\mathbb{Z}_2$-symmetric bistability at zero field ($H=0$). Under weak applied fields satisfying $|H| < |H_{\mathrm{sp}}^{\pm}|$, this double-well potential becomes tilted while maintaining both local minima.
By solving from $\partial_{\phi} f_4(\phi, H)=0$ and $\partial^2_{\phi} f_4(\phi, H)=0$, the spinodal field $H_{\mathrm{sp}}^{\pm} $ is obtained as
\begin{equation}
\label{eq:phi4-Hsp}
H_{\mathrm{sp}}^{\pm} = \pm H^* \equiv \pm \sqrt{(4/27) (-a_2^3/a_4)} ,
\end{equation}
at which the metastable state ceases to exist. This corresponds to a saddle-node bifurcation, where the metastable minimum merges with the unstable maximum at 
\begin{equation}
\label{eq:phi4-phisp}
\phi_{\mathrm{sp}}^{\pm} = \pm \phi^* \equiv \mp \sqrt{- a_2/(3a_4)}
\end{equation}
where the superscripts $\pm$ denote transitions toward positive and negative $\phi$, respectively. 
\begin{figure}[htbp]
    \centering
    \includegraphics[width=1\linewidth]{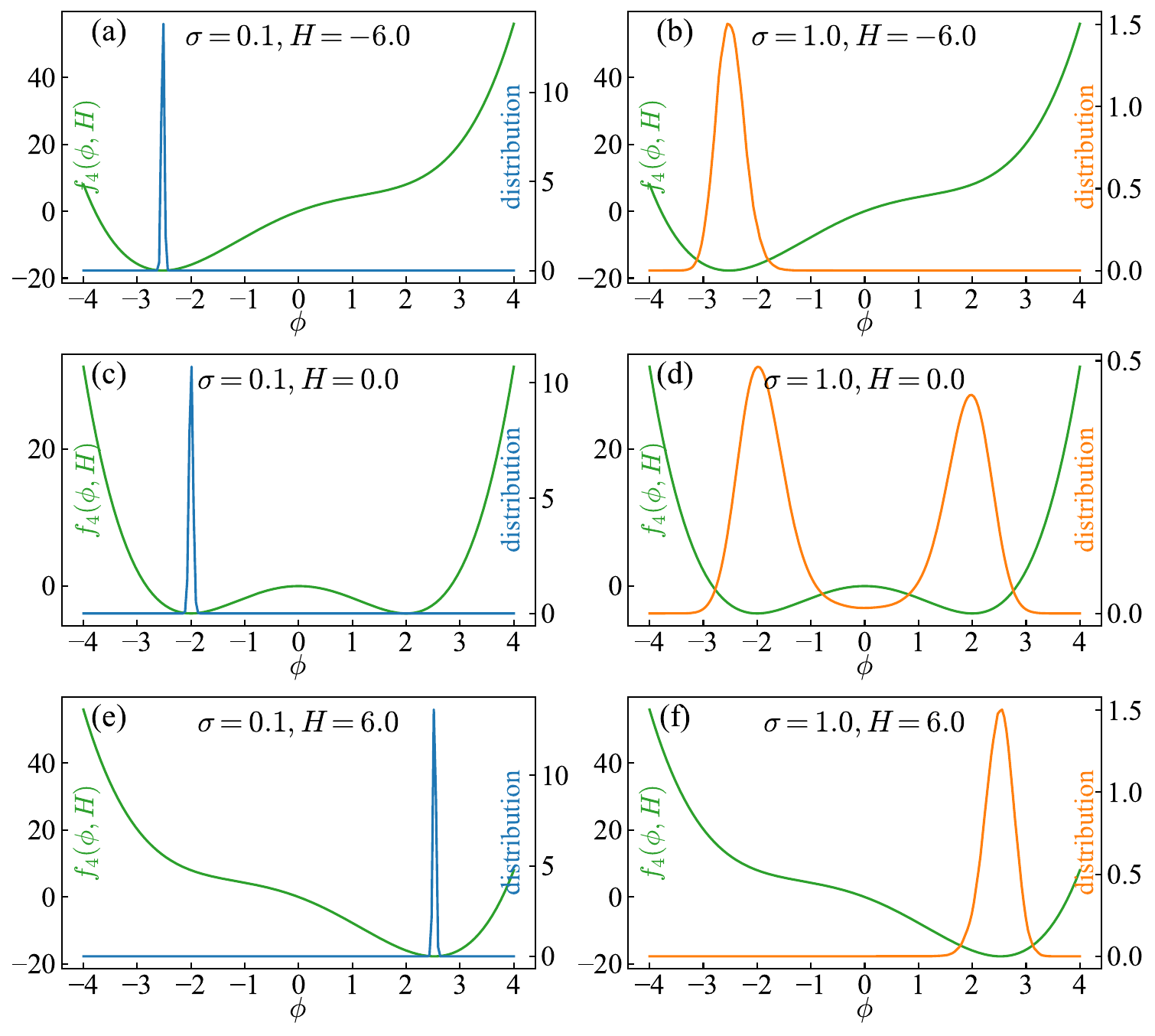}
    \caption{Potential curves and probability distributions of the stochastic $\phi^4$ model at different external field values in a driven process. The left and right columns correspond to $\sigma=0.1$ and $\sigma=1$, respectively. From top to bottom, $H=$ -6, 0, 6. In this plot, $a_2=-4$, $v_H=0.001$ and all quantities are nondimensionalized by $a_4$ and $\lambda$.}
    \label{SMfig:potential_curve}
\end{figure}
The dynamics of $\phi = \phi(t)$ under a time-dependent field $H = H(t)$ can be described by the Langevin equation~\cite{tuckerOnsetSuperconductivityOneDimensional1971,kunduDynamicHysteresisNoisy2023}
\begin{equation}
\label{eq:phi4-Langevin}
    \frac{\partial \phi}{\partial t} = -\lambda \frac{\partial f_4(\phi, H)}{\partial \phi} +\zeta(t).
\end{equation}
Here $\lambda$ sets a reference timescale and $\zeta(t)$ is a Gaussian white noise with amplitude $\sqrt{2\lambda\sigma^2}$ satisfying $\langle \zeta (t) \rangle=0$ and $\langle\zeta(t)\zeta(t^{\prime})\rangle=2\lambda\sigma^2\delta(t-t^{\prime})$, and $\sigma \ge 0$ denoting the noise strength.


Substituting the free energy expression Eq.~(\ref{eq:phi4-free-energy}) into the dynamic equation Eq.~(\ref{eq:phi4-Langevin}), the specific form of the dynamic equation reads,
\begin{equation}
    \frac{\partial \phi}{\partial t} = -\lambda a_2 \phi - \lambda a_4 \phi^3 + \lambda H + \zeta (t).
\end{equation}
The change of the order parameter over a small time interval $\delta t$ is given by the discretized form of the above equation~\cite{brankaAlgorithmsBrownianDynamics1998},
\begin{equation}
    \label{SMeq:random_increment_phi4}
    \delta\phi = -\lambda a_2 \phi\delta t - \lambda a_4 \phi^3 \delta t + \lambda H\delta t + \sqrt{2\lambda\sigma^2\delta t}W,
\end{equation}
where $W$ is a random variable drawn from a standard normal distribution. By simulating Eq.~\eqref{SMeq:random_increment_phi4}, we can obtain the evolution of $\phi(t)$ under varying field $H(t)$. Figure~\ref{SMfig:potential_curve} shows the potential curves and corresponding probability distributions in increasing-field processes for two different noise strengths.
From top to bottom, the field is strongly negative, zero, and strongly positive, respectively. For the smaller noise strength, the probability distribution remains highly concentrated around the initial minimum. The system retains its initial state well past the field where the two local minima become energetically symmetric. In contrast, for the larger noise strength, the wider probability distribution facilitates an earlier transition from the initial state to the final equilibrium state due to enhanced fluctuations.

\subsection{Hysteresis loops}
\begin{figure}[htbp]
    \centering
    \includegraphics[width=0.95\linewidth]{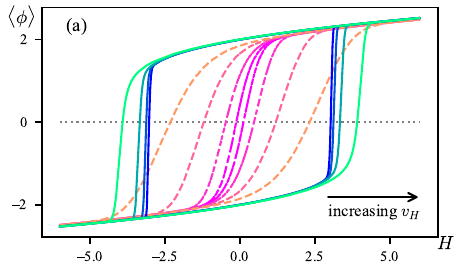}
    \includegraphics[width=0.95\linewidth]{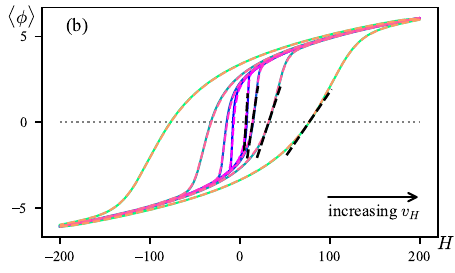}
    \caption{Hysteresis loops for $\sigma=0.1$ (solid lines) and $\sigma=1.0$ (dased lines). (a) Slow driving, $v_H=$0.004, 0.02, 0.1, 0.5 from the inside out. (b) Fast driving, $v_H=$8, 40, 200, 1000 from the inside out. $a_2=-4$ and all quantities nondimensionalized by $a_4$ and $\lambda$.}
    \label{SMfig:hysteresis_mono}
\end{figure}

We control the external field varying linearly between $\mp H_m$ at a rate $|dH/dt| = \lambda v_H$, where $H_m > 0$ is the amplitude.  With the system initially prepared in the equilibrium state corresponding to the strongly negative field $-H_m$, the averaged order parameter $\langle \phi \rangle$ evolves from a negative to a positive value and then returns to the negative one. As the external field changes, the system responds by adjusting $\langle \phi \rangle$ , which reflects the competition between the driving field and the system's internal dynamics. The point at which it crosses zero defines the coercivity, marking the field strength at which the system's response changes direction. 
 
 Figures~\ref{SMfig:hysteresis_mono}(a) and \ref{SMfig:hysteresis_mono}(b) show the hysteresis loops during relatively slow and fast driving processes, respectively. The coercivity, as reflected in the $\langle\phi\rangle-H$ curves, increases with the driving velocity. In the relatively slow driving regime, the curves exhibit distinct behaviors depending on the noise strength: for large noise strength, the transitions are smoother and the coercivity is smaller; for small noise strength, the transitions are sharper and the coercivity is larger. This difference arises from the noise-induced reduction in relaxation time. As the external field increases and the left potential well becomes metastable, the escape rate to the stable well is enhanced by larger noise strength~\cite{riskenFokkerPlanckEquationMethods1996}. As a result, the order parameter transitions earlier, at a smaller field value. In the relatively fast driving regime, the $\langle\phi\rangle - H$ curves for different noise strengths nearly overlap. This is because the driving velocity exceeds any intrinsic relaxation timescale of the system, regardless of the noise strength, leading to similar passive relaxation processes.

\section{Coercivity of the stochastic $\phi^4$ model}
\label{sec:coercivity_stochastic_phi4}  
Coercivity is a fundamental property of the hysteresis loop, especially for magnetic and ferroelectric materials, defining the external magnetic or electric field strength required to reduce a material's magnetization or polarization to zero. It indicates how well a material resists demagnetization or depolarization: high coercivity materials are "hard" and maintain their state effectively, ideal for permanent magnets in devices like electric motors~\cite{WOS:A1986A038800179}, while low coercivity materials are "soft" and easily switched, commonly used in transformers~\cite{8651370}. This property is vital for optimizing performance in technologies such as magnetic storage devices and sensors~\cite{pandey_l_2009,278712}. At a general level, in our current work, we define the coercivity $H_c\equiv H(\langle \phi \rangle=0)$ as a mesoscopic or macroscopic observable of the hysteresis phenomenon caused by microscopic dynamics. Clarifying the change of coercivity with the external driving is of great significance for theoretically quantifying the hysteresis phenomenon and engineering the performance of devices constructed with hysteretic materials in practical applications. Below, we provide a systematic exploration of the coercivity panorama proposed in the companion paper~\cite{CP}.

\subsection{Coercivity Panorama}

\begin{figure}[htbp]
    \centering
    \includegraphics[width=1\linewidth]{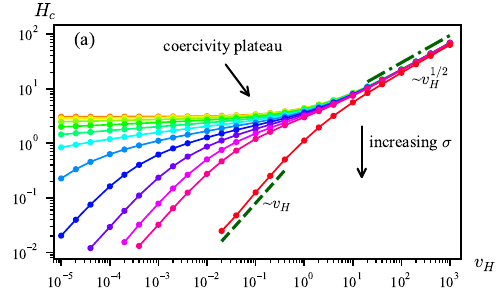}
    \includegraphics[width=1\linewidth]{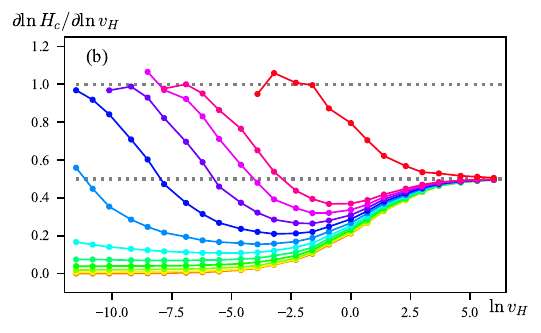}
    \caption{(a) Coercivity values and (b) their scaling exponents with respect to the driving rate within saturated hysteresis loops for different noise strengths. $\sigma=$0, 0.1, 0.2, 0.3, 0.4, 0.5, 0.6, 0.7, 0.8, 0.9, 1, 2 from top to bottom in (a) while from bottom to top in (b). In this plot, $a_2=-4$.}
    \label{SMfig:coercivity_vH}
\end{figure}

In [Fig.1e] of the companion paper~\cite{CP}, we presented a coercivity panorama. Here, a more comprehensive coercivity landscape is provided in Fig.~\ref{SMfig:coercivity_vH}. 
In Fig.~\ref{SMfig:coercivity_vH}(a), the curves from top to bottom correspond to increasing noise strength $\sigma$. For the largest noise strength, we include two reference lines in the fast- and slow-driving regimes, indicating power-law behaviors with exponents $1/2$ and $1$, respectively. For the largest noise strength [the topmost curve in Fig.~\ref{SMfig:coercivity_vH}(b)], the scaling exponent increases monotonically from $1/2$ to $1$ as the velocity decreases. However, for smaller noise strengths, the exponent first decreases before increasing toward $1$ in the very slow-driving regime. When the noise strength is very small, the exponent continues to decrease throughout the simulated velocity range, although it is expected to eventually increase at even slower driving. The decreasing behavior in the scaling exponent reflects a plateau-like region in the coercivity-velocity curves, where the coercivity changes very slowly with velocity. The plateau-like behavior of the coercivity has not been reported in previous studies. We interpret this plateau as a characteristic feature of FOPT, namely FOPT plateau. It becomes increasingly broad and pronounced as the noise strength decreases and, in the thermodynamic limit, it is expected to evolve into an asymptotic plateau, indicating a exact FOPT that is triggered at a finite coercivity value in the quasi-static limit. The specific FOPT plateau point is defined as the local minimum point shown in Fig.~\ref{SMfig:coercivity_vH}(b), while the plateau point for very small noise strength has not been achieved in the presented simulation timescale.

To explicitly understand the response of coercivity with respect to the driving rate, Eq.~\eqref{eq:phi4-Langevin} is equivalently expressed by the following Fokker-Planck equation,
\begin{equation}
\label{eq:random_FP_phi4}
    \frac{\partial P(\phi, t)}{\partial t} = \lambda \frac{\partial }{\partial \phi}\left[\frac{\partial f\left(\phi, H\right)}{\partial \phi} + \sigma^2 \frac{\partial}{\partial \phi}\right] P(\phi, t),
\end{equation}
where $P(\phi, t)$ is the probability density function. From the above equation, we obtain the evolution of $\langle \phi \rangle$ in time, 
\begin{equation}
\label{eq:random_avephi_phi4}
    v_H \frac{d\langle \phi \rangle}{dH} = - a_2 \langle \phi \rangle - a_4 \langle \phi^3 \rangle + H,
\end{equation}
where we have used the linear variation of $H(t)$, that is $H(t) = H_0 + \lambda v_H t$, and hence $dH = \lambda v_H dt$. Hereafter, the increasing branch of the field function is considered in analytical calculations, unless specifically stated otherwise. The second term on the right-hand side of the above equation may be expressed as
\begin{equation}
\label{eq:third_moment}
    \langle \phi^3 \rangle = \langle \phi \rangle^3 + 3 \mu_{2} \langle \phi \rangle + \mu_{3},
\end{equation}
where $\mu_{2} \equiv \langle (\phi - \langle \phi \rangle )^2 \rangle$, $\mu_3 \equiv \langle (\phi - \langle \phi \rangle)^3 \rangle$ are the second and third central moments of the random variable $\phi$, respectively. In particular, $\mu_3$ is associated with the skewness of the distribution of $\phi$. The coercivity is expressed as
\begin{equation}
\label{eq:random_coercivity_phi4}
    H_c = v_H \eval{\frac{d\langle \phi \rangle}{d H}}_{\langle \phi \rangle = 0} + a_4 \eval{\mu_3}_{\langle \phi \rangle = 0}.
\end{equation}

\subsection{$H_c \sim v_H$ scaling in the near equilibrium regime}

Linear scaling is observed in the near-equilibrium regime. In the equilibrium limit, the order parameter is distributed symmetrically in the double-well potential landscape, and therefore $\langle \phi \rangle$ becomes zero at $H = 0$, indicating that the coercivity vanishes in this limit. Due to this symmetric distribution, the skewness of the equilibrium distribution, represented by the third central moment $\mu_3$, also vanishes. In one word, both the equilibrium-limit values of $H_c$ and $\mu_3$ at $\langle \phi \rangle = 0$ are zero.

To quantify deviations from equilibrium in the near-equilibrium regime, the key quantities, $H_c$, $\eval{\mu_3}_{\langle \phi \rangle = 0}$, and $\eval{\left( d \langle \phi \rangle/d H \right)}_{\langle \phi \rangle = 0}$, in Eq.~\eqref{eq:random_coercivity_phi4} can be respectively expanded in power series around $v_H=0$,

\begin{subequations}
    \begin{align}
        & H_c = h_1 v_H^{\alpha} + h_2 v_{H}^{2\alpha} + \dots \\
        & \mu_3 = u_1 v_H^{\beta} + u_2 v_{H}^{2\beta} + \dots \\ 
        & \eval{\frac{d\langle \phi \rangle}{d H}}_{\langle \phi \rangle = 0} = \eval{\frac{d\langle \phi \rangle_{\mathrm{eq}}}{d H}}_{\langle \phi \rangle_{\mathrm{eq}} = 0} + g_1 v_H^{\gamma} + g_2 v_{H}^{2\gamma} + \dots, \label{SMeq:linear_scaling_expansion_phiHgrad}
    \end{align}
\end{subequations}
where the subscript $\mathrm{eq}$ denotes equilibrium limit. Substituting the above series into Eq.~\eqref{eq:random_coercivity_phi4} and retaining of the lowest-order corrections in each term, we find that $\alpha = \beta = 1$ in order to balance the powers of $v_H$. Therefore, the near-equilibrium correction of the coercivity follows as
\begin{equation}
    H_c\propto v_H.
\end{equation}
We emphasize that the above analysis applies to finite noise strength, as $\eval{\left( d \langle \phi \rangle/d H \right)}_{\langle \phi \rangle = 0}$ diverges in the equilibrium limit when the noise vanishes, and thus cannot be expanded in a power series like Eq.~\eqref{SMeq:linear_scaling_expansion_phiHgrad} near equilibrium. 

\subsection{$H_c \sim v_H^{1/2}$ scaling in fast-driving regime}

In the fast-driving regime, coercivity shows no significant dependence on noise strengths. This is because the external field changes so rapidly that the system remains effectively frozen before the potential landscape becomes monostable. In this regime, the dynamics are dominated by deterministic forces, while fluctuations, which are crucial for transitions between metastable states, play a negligible role. Consequently, the coercivity behavior can be understood by analyzing the relaxation dynamics toward the monostable state in systems with vanishing fluctuations under a time-dependent strong field. 

For a varying external field, the position of the monostable state, that is the equilibrium state, in the $\phi^4$ model also changes accordingly. When the magnitude of the field is sufficiently large, specifically for $|H| \gg \sqrt{|a_2^3 / a_4|}$, the equilibrium state can be approximated as follows
\begin{equation}
    \phi_{\mathrm{eq}} = \left( \frac{H}{a_4} \right)^{1/3},
\end{equation}
In deriving the above expression, the $a_2 \phi$ term in $\partial f(\phi, H) / \partial \phi$ has been neglected, as it becomes negligible compared to the $a_4 \phi^3$ term when the external field $H$ is sufficiently large.

The relaxation dynamics near the equilibrium state can be approximated by the following linearized equation,
\begin{equation}
    \frac{\partial \langle \phi \rangle}{\partial t} \approx - \lambda \eval{\frac{\partial^2 f(\phi, H)}{\partial \phi^2}}_{\phi_{\mathrm{eq}}} \left(\langle \phi \rangle - \phi_{\mathrm{eq}}\right),
\end{equation}
from which we obtain that the characteristic relaxation time to the equilibrium state is
\begin{equation}
\begin{aligned}
    \tau_{\mathrm{relax}} =& \frac{1}{\lambda \eval{(\partial^2f / \partial \phi^2)}_{\mathrm{eq}} } = \frac{1}{\lambda (a_2 + 3 a_4 \phi_{\mathrm{eq}}^2)}\\
    \approx & \frac{1}{3\lambda a_4 \phi_{\mathrm{eq}}^2} \\
    \approx & \frac{1}{3 \lambda a_4^{1/3} H^{2/3}},
\end{aligned}
\end{equation}
where we have used the expression of $\phi_{\mathrm{eq}}$.

With an increasing external field, the equilibrium state moves with the rate
\begin{equation}
    r_{\phi_{\mathrm{eq}}} \equiv \frac{d \phi_{\mathrm{eq}}}{dH} v_H \approx \frac{1}{3} \frac{v_H}{a_4 ^{1/3} H^{2/3}}.
\end{equation}
The characteristic rate of relaxation is
\begin{equation}
    r_{\mathrm{relax}} \equiv \frac{1}{\tau_{\mathrm{relax}}} \approx 3\lambda a_4^{1/3} H^{2/3}.
\end{equation}
Since the characteristic rate of relaxation increases and the moving rate of the equilibrium state decreases with the external field for fixed $v_H$, we identify that the equation
\begin{equation}
    r_{\mathrm{relax}} \simeq r_{\phi_{\mathrm{eq}}}
\end{equation}
represents a turning point of the field that the relaxation of the system becomes able to catch up the variation of the equilibrium state and we define this turning point as $H_{\mathrm{satur}}$. By substituting the expressions of $r_{\mathrm{relax}}$ and $r_{\phi_{\mathrm{eq}}}$ into the above equation, we obtain that $H_{\mathrm{satur}}$ scales with $v_H$ as follows,
\begin{equation}
    H_{\mathrm{satur}} \sim v_H^{3/4}.
\end{equation}

\begin{figure}[htbp]
    \centering
    \includegraphics[width=1\linewidth]{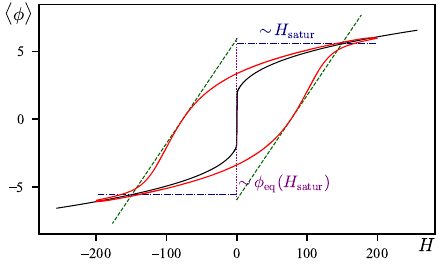}
    \caption{Geometric method to explain the scaling exponent in the fast driving regime. $\sigma=0.7, v_H=1000$ are used to generate the hysteresis loop while the central transition curve represents the equilibrium limit.}
    \label{SMfig:hysteresis_fast_driving_triangle}
\end{figure}

To associate $H_{\mathrm{satur}}$ with the interested coercivity, 
we introduce a reference right triangle on the hysteresis curve. The hypotenuse of the triangle is defined by the line passing through the coercivity point with the same slope as the $\langle \phi \rangle - H$ curve at this point, and it intersects the curve again at its terminal point. The vertical leg of the triangle is defined to pass through the equilibrium curve, and the horizontal leg is determined accordingly.

Since the averaged order parameter value approaches the equilibrium value at the right terminal point of the horizontal leg, the length of the horizontal leg is approximated by the magnitude of $H_{\mathrm{satur}}$. Accordingly, the length of the vertical leg is approximated by $\phi_{\mathrm{eq}}(H_{\mathrm{satur}})$. Then the slope of the hypotenuse is approximated by $\phi_{\mathrm{eq}}(H_{\mathrm{satur}}) / H_{\mathrm{satur}}$, which gives the following scaling relation with respect to $v_H$,
\begin{equation}
    \eval{\frac{d \langle \phi\rangle}{dH}}_{\langle \phi \rangle =0} \sim H_{\mathrm{satur}}^{-2/3} \sim v_H^{-1/2}.
\end{equation}

Furthermore, in the fast driving regime, the skewness of $\phi$ is negligible in Eq.~\eqref{eq:random_coercivity_phi4}, and hence the coercivity is approximated by
\begin{equation}
\label{eq:random_coercivity_phi4_fast}
    H_c = v_H \eval{\frac{d\langle \phi \rangle}{d H}}_{\langle \phi \rangle = 0},
\end{equation}
which is verified in Fig.~\ref{SMfig:hysteresis_mono}(b), where the tilted dashed line is plotted with the slope $H_c/v_H$.  Hence, the scaling behavior of the coercivity with respect to driving velocity in the fast driving regime is
\begin{equation}
    H_c \sim v_H^{1/2}.
\end{equation}

\subsection{Finite-size scaling of the FOPT plateau}

In this subsection, we investigate how finite-size effects influence the coercivity plateau as characteristic of FOPT. By expanding the potential near the spinodal point in the deterministic limit, we derive how the plateau height and width scale with noise strength.

Since the plateau is associated with FOPT, we will expand the quartic potential near the transition point defined in the thermodynamic limit, retaining terms up to the cubic order of the order parameter~\cite{jungScalingLawDynamical1990,zhongCompleteUniversalScaling2024}. This is justified by the fact that the order parameter undergoes an asymmetric discontinuous flip during FOPT. The transition point of the system in the thermodynamic limit is exactly the spinodal point $(H^*,\phi^*)$ of the potential $f(\phi,H)$ given by Eqs.~\eqref{eq:phi4-Hsp} and \eqref{eq:phi4-phisp}. Defining $h \equiv H - H^*$ and $\varphi \equiv \phi - \phi^*$, the original quartic potential~\eqref{eq:phi4-free-energy} is expanded as 
\begin{equation}
\label{eq:phi4-free-energy-phi3}
    f(\varphi, h) = \frac{1}{3} a_3 \varphi^3 - h \varphi,
\end{equation}
where $a_3 = - \sqrt{-3 a_2 a_4}$. Substitute the above cubic potential into the governing Langevin equation~\eqref{eq:phi4-Langevin} and then one obtains the corresponding discretized equation as
\begin{equation}
\label{eq:random_increment_phi3}
    \delta \varphi = - a_3 \varphi^2 v_H^{-1} \delta h + h v_H^{-1} \delta h + \sqrt{2 \sigma^2 v_H^{-1} \delta h}W,
\end{equation}
where we have used the relation $\delta h = \lambda v_H \delta t$ for linearly increasing field. 
In Ref.~\cite{zhongCompleteUniversalScaling2024}, Zhong applies scaling transformations to the dynamic equation governed by the cubic potential, based on renormalization-group theory, and derives complete scaling relations for all system parameters and variables with respect to the driving velocity. In the present work, to investigate the finite-size effect on the coercivity plateau, we perform analogous scaling transformations with respect to the noise strength. Using $\Sigma$ as the scaling factor to scale the noise strength from $\sigma$ to $\sigma^{\prime} = \sigma \Sigma^{-1}$, the quantity $\mathcal{O}$ becomes $\mathcal{O}^{\prime} = \mathcal{O} \Sigma^{-[\mathcal{O}]}$, where $[\mathcal{O}]$ is the scaling dimension of $\mathcal{O}$. Performing scaling transformations on Eq.~\eqref{eq:random_increment_phi3}, one obtains
\begin{equation}
    \begin{aligned}
        \delta \varphi^{\prime} \Sigma^{[\varphi]} =& - a_3^{\prime} \varphi^{\prime 2} v_{H}^{\prime -1} \delta h^{\prime} \Sigma^{[a_3] + 2 [\varphi] - [v_H] + [h]}\\
        & + h^{\prime} v_{H}^{\prime -1} \delta h^{\prime} \Sigma^{2 [h] - [v_H]} \\
        & + \sqrt{2\sigma^{\prime 2} v_H^{\prime -1} \delta h^{\prime}} \Sigma^{1 - [v_H] / 2 + [h] / 2}.
    \end{aligned}
\end{equation}
To keep the dynamic equation invariant after scaling transformation, the scaling dimensions of each term in the equation should balance, that is
\begin{equation}
\label{SMeq:scaling_balance_phi3}
    [\varphi] = [a_3] + 2 [\varphi] - [v_H] + [h] = 2 [h] - [v_H] = 1 - [v_H] / 2 + [h] / 2.
\end{equation}
According to renormalization-group theory, the scaling dimension of the coefficient in the highest-order term is set to be zero~\cite{zhongCompleteUniversalScaling2024}, that is $[a_3] = 0$. Substituting $[a_3] = 0$ into Eq.~\eqref{SMeq:scaling_balance_phi3}, we obtain that
\begin{equation}
    [h] = \frac{4}{3}, \quad [\varphi] = \frac{2}{3}, \quad [v_H] = 2.
\end{equation}
The above scaling dimensions reveal that the scaling exponents of $h$, $\varphi$ and $v_H$ with respect to $\sigma$ are respectively $4/3$, $2/3$ and $2$. These scaling exponents apply to cases where the hysteresis curves are adjacent to the quasi-static phase-transition curve in the deterministic limit. For finite $\sigma$, this corresponds to the FOPT plateau regime. Although these scaling relations are directly applicable to turning points analogous to the spinodal point within the hysteresis curves, it is verified that coercivity points also exhibit these scaling relations due to the small difference between turning points and coercivity points in abrupt transition processes.
Therefore, we conclude the scaling exponents as
\begin{equation}
    | H_P - H^* | \sim \sigma ^{4/3} \quad \text{at} \quad v_{P} \sim \sigma^2,
\end{equation}
where $H_P$ is the height of the coercivity plateau, $| H_P - H^* |$ is the difference in the height between systems with finite and vanishing noise strengths and $v_{P}$ is the reference driving rate at which the coercivity plateau is observed. Considering that the plateau height decreases with increasing $\sigma$, as shown in Fig.~\ref{SMfig:coercivity_vH}(a), we rewrite the above relations as
\begin{equation}
\label{eq:phi4-finite-size-scaling-plateau}
    H^* - H_{P} \sim \sigma^{4/3} \quad \text{at} \quad v_{P} \sim \sigma^2.
\end{equation}
This finite-size scaling is numerically validated in [Fig. 2(a)] of the companion paper~\cite{CP}.

\subsection{$H_c-H_P\sim (v_H-v_P)^{2/3}$ scaling in the post-plateau regime}
Building on the characterization of the coercivity plateau, we now analyze the finite-time scaling in the post-plateau. According to Eq.~\eqref{eq:random_avephi_phi4}, for a given initial condition, the hysteresis curve under linearly varying field is uniquely determined by the driving rate $v_H$. Considering the specific hysteresis curve corresponding to $v_P(\sigma)$ for some $\sigma$, denote the turning point that triggers a rapid transition within this curve as $(H_P^*, \phi_P^*)$. Expanding the free energy around this turning point and retaining to the third order term, we have
\begin{equation}
\label{SMeq:phi4-plateau-phi3-energy}
    f_3(\varphi_P,h_P)= b_0 + b_1 \varphi_P + \frac{1}{2}b_2 \varphi_P^2 + \frac{1}{3}b_3 \varphi_P^3 - \phi_P^* h_P - h_P\varphi_P,
\end{equation}
where $\varphi_P\equiv \phi-\phi_P^*$, $h_P\equiv H-H_P^*$, and
\begin{equation*}
    \begin{aligned}
        & b_0 = \frac{1}{4}a_4\phi_P^{*4}+\frac{1}{2}a_2\phi_P^{*2}-H_P^*\phi_P^*,\\
        & b_1 = a_4\phi_P^{*3}+a_2\phi_P^*-H_P^*,\\
        & b_2 = 3a_4\phi_P^{*2} + a_2,\\
        & b_3 = 3a_4\phi_P^*.
    \end{aligned}
\end{equation*}
Substituting Eq.~\eqref{SMeq:phi4-plateau-phi3-energy} into the governing Fokker-Planck equation~\eqref{eq:random_FP_phi4},
\begin{equation}
\begin{aligned}
\frac{\partial P(\varphi_P,t)}{\partial t} =\, & 
\lambda\frac{\partial}{\partial\varphi_P}\left[\left(b_1 + b_2\varphi_P + b_3\varphi_P^2 - h_P\right)P(\varphi_P,t)\right] \\
& + \lambda\sigma^2\frac{\partial^2 P(\varphi_P,t)}{\partial\varphi_P^2}.
\end{aligned}
\end{equation}
Multiplying both sides of the above equation by $\varphi_P$ and integrating over $\varphi_P$, one obtains the averaged equation
\begin{equation}
    v_H\frac{d\langle \varphi_P \rangle}{dh_P} = -b_1 - b_2 \langle \varphi_P \rangle - b_3 \langle \varphi_P^2 \rangle + h_P,
\end{equation}
where the relation $dh_P=\lambda v_H dt$ has been used. Evaluating the above equation at $\langle \varphi_P \rangle=0$, the result reads
\begin{equation}
\label{SMeq:phi4-plateau-phi3-slope}
    v_H \eval{\frac{d\langle \varphi_P \rangle}{dh_P}}_{\langle \varphi_P \rangle=0} = -b_1 - b_3 \eval{\langle\varphi_P^2\rangle}_{\langle\varphi_P\rangle=0} + \eval{h_P}_{\langle\varphi_P\rangle=0}.
\end{equation}
For $v_H=v_P$, the above relation becomes
\begin{equation}
    v_P \eval{\frac{d\langle \varphi_P \rangle}{dh_P}}_{\langle \varphi_P \rangle=0, v_P} = -b_1 - b_3 \eval{\langle\varphi_P^2\rangle}_{\langle\varphi_P\rangle=0, v_P}.
\end{equation}
Defining $\nu_h\equiv v_H-v_P$ and expanding Eq.~\eqref{SMeq:phi4-plateau-phi3-slope} around $\nu_h=0$, we obtain
\begin{equation}
\label{CPeq:phi4-plateau-phi3-slope-expansion}
    \nu_h s = -b_3 \mu + \eval{h_P}_{\langle\varphi_P\rangle=0},
\end{equation}
where
\begin{equation}
\begin{aligned}
s &\equiv \left. \frac{d\langle \varphi_P \rangle}{dh_P} \right|_{\langle \varphi_P \rangle=0} - \left. \frac{d\langle \varphi_P \rangle}{dh_P} \right|_{\langle \varphi_P \rangle=0, v_P}, \\
\mu &\equiv \left. \langle \varphi_P^2 \rangle \right|_{\langle \varphi_P \rangle=0} - \left. \langle \varphi_P^2 \rangle \right|_{\langle \varphi_P \rangle=0, v_P}.
\end{aligned}
\end{equation}
Performing scaling transformations on Eq.~\eqref{CPeq:phi4-plateau-phi3-slope-expansion} with respect to $\nu_h$, which gives $[\nu_h]=1$, $[s] \sim [\varphi_P]-[h_P]$, $[\mu] \sim 2[\varphi_P]$, and setting $[b_3]=0$, the resulting balanced scaling relation is $\eval{h_P}_{\langle\varphi_P\rangle=0}\sim \nu_h^{2/3}$. Due to the closeness between the turning point and the coercivity point for hysteresis loops with a feature of first-order phase transition, the coercivity value also adheres to this scaling, that is 
\begin{equation}
    H_c-H_P\sim (v_H-v_P)^{2/3}.
\end{equation}

\section{Coercivity panorama of the Curie-Weiss model}
\label{sec:coercivity_panorama_CW}  
In this section, we apply the established coercivity panorama to study a specific microscopic model, the Curie-Weiss model~\cite{friedliStatisticalMechanicsLattice2017,meibohmFiniteTimeDynamicalPhase2022,wuErgodicityBreakingScaling2025,fioriSpecificHeatDriven2025}. We will extend the discussion on the stochastic $\phi^4$ model to systematically study the finite-time and finite-size scaling of the coercivity in the magnetic hysteresis of the Curie-Weiss model. Furthermore, the universal scalings and model-dependent scalings in the coercivity panorama will be clarified.

 
\subsection{Dynamics and Hysteresis}
\label{sec:CW_Dynamics}

The Curie-Weiss model is a paradigmatic mean-field model for ferromagnetic systems, capturing phase transitions through infinite-range interactions. Consider a system of \(N\) Ising spins \(\sigma_i = \pm 1\), labeled $i=1,...,N$, where each spin interacts as the total magnetization with all others via a ferromagnetic coupling $J/2N$, and define the total magnetization $M = \sum_{i=1}^N \sigma_i$. The system is immersed in a thermal bath at inverse temperature $\beta=1/(k_BT)$ and driven by an external time-dependent magnetic field $H(t)$.  The Hamiltonian is given by~\cite{friedliStatisticalMechanicsLattice2017,wuErgodicityBreakingScaling2025}:
\begin{equation}
\mathcal{H} = -\frac{J}{2N} M^2 - H(t) M,    
\end{equation}
When an arbitrary spin flips from $\mp1$ to $\pm1$, the total magnetization will change by two units, i.e., $M\to M_{\pm}\equiv M \pm 2$. The probability \(P(M, t)\) of observed total magnetization $M$ is governed by the master equation
\begin{equation}
\label{SMeq:dPdtCW}
\begin{aligned}
\frac{dP(M, t)}{dt} = 
\sum_{\eta=\pm} \Big[ & W_\eta(M-\eta, H)P(M-\eta, t) \\
& - W_\eta(M, H)P(M, t) \Big],
\end{aligned}
\end{equation}
with the transition rates for $M\to M_{\pm}\equiv M \pm 2$,
\begin{equation}
\label{SMeq:CW-transition-rate}
W_\pm(M, H) = \frac{N \mp M}{2\tau_0} \exp\left[\pm \beta \left(J\frac{M \pm 1}{N} + H\right)\right]
\end{equation}
and the microscopic relaxation time $\tau_0$.
These transition rates satisfy the detailed balance condition
\begin{equation}
W_\pm(M, H) P_\text{eq}(M, H) = W_\mp(M_\pm, H) P_\text{eq}(M_\pm, H),
\end{equation}
where \( P_\text{eq}(M, H) \propto \exp\left[ -\beta \mathcal{F}_H(M) \right] \) is the equilibrium distribution with the free
energy
\begin{equation}
\mathcal{F}_H(M) = -\frac{J}{2N}M^2 - HM - \frac{1}{\beta} \ln \Omega(M),
\end{equation}
and $ \Omega(M) = N!/([(N + M)/2]![(N - M)/2]!)$ is the microscopic  degeneracy of magnetization \( M \).

\begin{figure}
    \centering
    \includegraphics[width=1\linewidth]{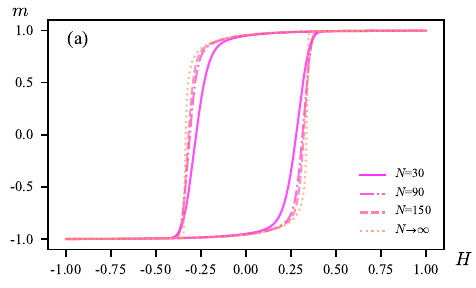}
    \includegraphics[width=1\linewidth]{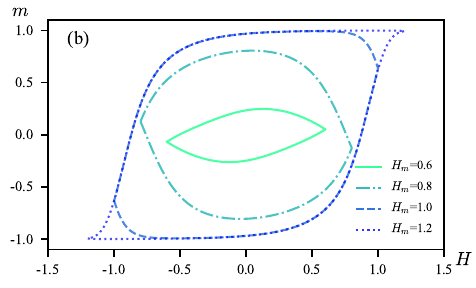}
    \caption{Dynamic hysteresis loops of the Curie–Weiss model under periodic driving for (a) different spin number $N$ at fixed amplitude $H_m=1$;(b) different driving amplitude $H_m$ with $N=30$. In this plot, the parameters $\tau_0=2$, $\beta=2$, and $J=1$ are used. Unless otherwise stated, the parameters used in the other figures of this section are the same as those here.}
    \label{fig:CW_coercivity_vH}
\end{figure}

By multiplying both sides of Eq.~\eqref{SMeq:dPdtCW} by $M$ and sum over all possible states, the time evolution of the average magnetization $\langle M \rangle$ is given by
\begin{equation}
\begin{split}
\frac{d\langle M \rangle}{dt} =\, & \sum_{M} \sum_{\eta = \pm} \bigg[ MW_\eta(M - \eta, H) P(M - \eta, t) \\
& - MW_\eta(M, H) P(M, t) \bigg],
\end{split}
\end{equation}
which, in the thermodynamic limit of $N \to \infty$, becomes the deterministic evolution equation of $m\equiv M/N$~\cite{meibohmFiniteTimeDynamicalPhase2022,wuErgodicityBreakingScaling2025}

\begin{equation}
\label{SMeq:dmdt}
\begin{aligned}
\frac{dm}{dt} =\, & \frac{2}{\tau_0} \Big[ \sinh{(\beta Jm + \beta H)} \\
& - m \cosh{(\beta Jm + \beta H)} \Big] \equiv g(m, H).
\end{aligned}
\end{equation}
By solving this equation with a linear driving protocol with an amplitude $H_m$ and a rate of $dH/dt = v_H$, we obtain the hysteresis loop for the infinite system ($N\to\infty$) as the solid curve in Fig.~\ref{fig:CW_coercivity_vH}(a) with $v_H=0.001$. In this example, we use  $\tau_0=2$, $\beta=2$, and $J=1$. Unless otherwise stated, the parameters used in the other figures of this section are the same as those here.

To illustrate the finite-size effect on hysteresis loop, we numerically solve the master equation Eq.~\eqref{SMeq:dPdtCW} to obtain the probability distribution $P(M,t)$, and compute the per-site average magnetization  
$\langle m\rangle = [\sum_M MP(M,t)]/N,$ for $N=30,90,150$. As shown in Fig.~\ref{fig:CW_coercivity_vH}(a), increasing the system size $N$ sharpens the loops-thermal fluctuations are progressively suppressed and the curves approach the deterministic, almost rectangular shape expected in the thermodynamic limit. Besides, in Fig.~\ref{fig:CW_coercivity_vH}(b), with fixed $N=30$ and varied field amplitude $H_m$, larger $H_m$ results in wider loops and higher coercivity while small $H_m$ leads to unsaturated, narrow and flat hysteresis loops. This trend arises because only strong fields enable spins to overcome energy barriers for complete state transitions.

\begin{figure}[htbp]
    \centering
    \includegraphics[width=1\linewidth]{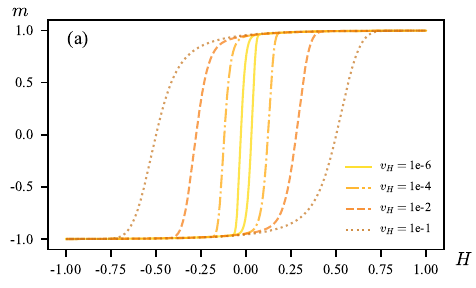}
    \includegraphics[width=1\linewidth]{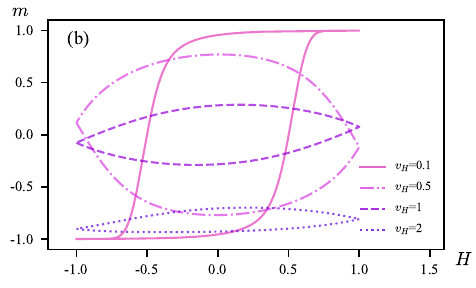}
    \includegraphics[width=1\linewidth]{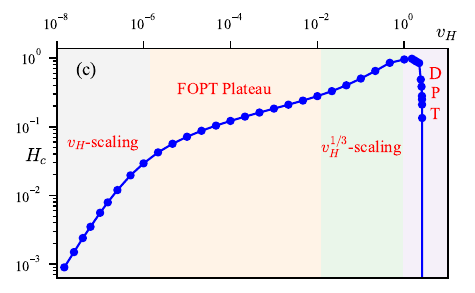}
    \caption{Hysteresis loops and coercivity panorama of the Curie–Weiss model with $N=30$ and $H_m=1$.  
    (a) Slow‐driving loops go through the the near-equilibrium regime, the FOPT plateau and the post-plateau.
    (b) Fast‐driving loops transition from the post-plateau to the DPT regime.
    (c) Coercivity $H_c$ versus driving rate $v_H$ on a log–log scale.}
  \label{fig:CW_panorama}
\end{figure}

Next, we will mainly focus on how the driving rate sculpts both loop morphology and coercivity at a fixed amplitude. In the slow‐drive regime Fig.~\ref{fig:CW_panorama}(a), the curves remain nearly symmetric, opening gradually as $v_H$ increases. As we move into the fast‐drive regime Fig.~\ref{fig:CW_panorama}(b), hysteresis loops become increasingly tilted and asymmetric, marking the onset of dynamic phase transitions (DPT). Fig.~\ref{fig:CW_panorama}(c) plots the coercivity panorama as a function of $v_H$, linking loop morphology across all rates. This panorama strikingly reveals four distinct dynamic regimes,  analogous to those of the stochastic $\phi_4$ model in~\cite{CP}: (i) a near-equilibrium regime where coercivity scales linearly with $v_H$; (ii) a plateau associated with FOPT; (iii) a post-plateau power-law scaling regime ; and (iv)a DPT regime with coercivity rapidly diminishing at high driving rates. The $v_H$-scaling in the near-equilibrium regime and the finite-size scaling relations associated with the FOPT plateau are universal across both the stochastic $\phi^4$ and Curie-Weiss models. It should be noted that the post-plateau power-law exhibits a $v_H^{1/3}$ scaling rather than the $v_H^{1/2}$ scaling obtained in the stochastic $\phi_4$ model, which will be calculated in detail in the next section.

\begin{figure}[htbp]
  \centering
  \includegraphics[width=\linewidth]{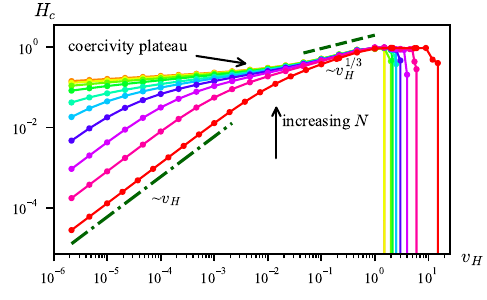}
  \caption{Coercivity $H_c$ as a function of  driving rate $v_H$ for system sizes at $N=5,10,15,20,25,30,40,50,60,70$ on a log–log scale. Other parameters as in
  Fig.~\ref{fig:CW_panorama}(c).}
  \label{fig:CW_coercivity_lines}
\end{figure}

Building on the single-system-size $N=30$ analysis above, Figure~\ref{fig:CW_coercivity_lines} extend this investigation by incorporating multiple system sizes $N$ under the same parameter settings, thereby illustrating how system size controls the coercivity panorama. At the small rate range of the figure, some curves of small systems from $5$ to $30$, collapse onto the near-equilibrium linear regime ($H_c\propto v_H$). For larger systems, a smaller driving speed (less than $10^{-6}$) is required to observe this scaling. As $v_H$ increases, a plateau appears where $H_c$ remains essentially constant; the width of this plateau grows with $N$, and its onset shifts to lower $v_H$ as finite-size effect diminishes. Immediately following, at the highest rates, all curves converge onto the fast-drive asymptote ($v_H^{1/3}$). Ultimately, all system sizes $N$ transition to the DPT regime. The larger the system size, the faster the coercivity decreases to zero. The behaviors described above are obtained under the condition of controlled field amplitude $H_m$, where the hysteresis loops do not necessarily reach full saturation. In contrast, if the external field amplitude $H_m$ is not constrained and the hysteresis loops are driven into full saturation, additional scaling behaviors emerge at high driving rate, which will be analyzed later.

\subsection{Finite-time scalings}
\label{sec:finite_time_scalings_CW} 
To further characterize the dynamic begavior observed in the hysteresis loops and their dependence on system size and driving rate, we now turn to dissect the finite-time scaling of coercivity in the Curie-Weiss model.

\subsubsection{$H_c-H^* \sim v_H^{2/3}$}
\label{sec:2/3-scaling}
During the dynamic process described by Eq.~\eqref{SMeq:dmdt} for infinite-size system, the system evolves along trajectories that may adhere to local equilibrium in either metastable or globally stable states, until a sudden transition occurs into a distinct state. This transition, contingent on the system’s history, exemplifies a finite-time first-order phase transition, driven by the collapse of the metastable state. The turning points $(\pm H^*,\pm m^*)$ are solved from $g(m,H)=0$ and $\partial_mg(m,H)=0$ as 
\begin{equation}
\label{SMeq:CW-spinodal-point}
\begin{aligned}
m^{*} &= -\sqrt{1 - \frac{1}{\beta J}}, \\
H^{*} &= \frac{1}{\beta}\left[\sqrt{\beta J(\beta J - 1)} - \operatorname{arctanh}\left(\sqrt{\frac{\beta J - 1}{\beta J}}\right)\right]
\end{aligned}
\end{equation}

\begin{figure}
    \centering
    \includegraphics[width=1\columnwidth]{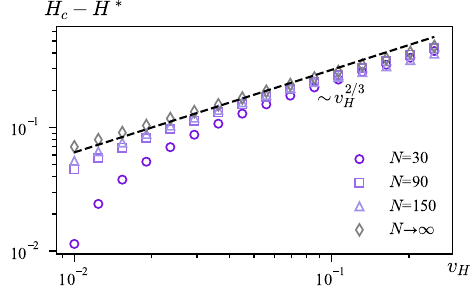} 
    \caption{Coercivity delay behavior of the Curie-Weiss model as a function of driving rate $v_H$. Purple dots are obtained by numerically solving the master equation Eq.~(\ref{SMeq:dPdtCW}) in finite-size systems. Gray dots are obtained by numerically solving the deterministic equation Eq.~(\ref{SMeq:dmdt}) for infinite system.} 
    
    \label{SMfig:CW-coercivity delay behavior-figure}
\end{figure}

Notably, when the sudden change occurs at a finite rate, the system's dynamics lags behind the quasistatic evolution, giving rise to an excess work between dynamic and quasistatic hysteresis loops. Such an additional energy cost is obtained in Ref.~\cite{wuErgodicityBreakingScaling2025} as 
\begin{equation}
w_{\text{ex}} \approx- \frac{A_1 ( 1 + \sqrt{1 - \frac{1}{\beta J}}) }{[4\sqrt{\beta J (\beta J - 1)} \frac{\beta} {\tau_0^2}]^{1/3}} v^{2/3},
\end{equation}
with the first zero point of the Airy function $A_1 \approx -2.338$~\cite{jungScalingLawDynamical1990}, which extends to various systems undergoing first-order phase transitions, including open quantum systems~\cite{drummond1980quantum,alodjants2024superradiant,zhang2021driven} and chemical reaction networks~\cite{nguyen2020exponential,schlogl1972chemical,ge2011non}. The results above are due to neglecting the fluctuations that vanish in infinitely large systems. However, when the driving rate is slow but not slower than the system's relaxation rate in large systems, nonergodicity is prominent. Consequently, the quasistatic work can be approximated by the deterministic limit and the average excess work scales as $w_{\text{ex}} \sim v^{2/3}$. We note that, for systems not extremely small with a slow driving rate, the excess work can be approximated by the enclosed rectangular area bounded by the upper and lower bounds of magnetization $m$ and the interval between $H_c$ and $H^*$,
\begin{equation}
w_{\text{ex}} \approx \Delta m(H_c-H^*),
\end{equation}
where the magnetization difference across the transition $\Delta m\approx 2$ is not related to $v_H$. This leads to the scaling relation
\begin{equation}
H_c-H^*\sim v_H^{2/3},
\end{equation}
which is numerically verified in the Fig. \ref{SMfig:CW-coercivity delay behavior-figure}. It can be found that the $2/3$ scaling of the coercivity delay behavioris locally observed in finite-size systems: the smaller the system, the onset of deviation shifts to higher $v_H$ more significantly, and the deviation becomes more pronounced. This behavior is closely related to the finite-size effects on the scaling of irreversible work discussed in~\cite{wuErgodicityBreakingScaling2025}, which finds that as the system size decreases, the scaling of the average excess work transitions from $v_H^{2/3}$ to $v_H$—an analogous transition driven by the same underlying physical mechanism. The phenomenon stems from the recovery of ergodicity and sufficiently strong thermal fluctuations capable of overcoming the energy barrier between the two minima,originating from the interplay between the driving rate and the relaxation rate dictated by the system size. Both observations indicate that finite-size effects fundamentally alter both the scaling behavior of the average excess work and coercivity, as well as the dynamic behavior of energy dissipation under external driving.

\subsubsection{$H_c \sim v_H^{1/3}$}
While the potential landscape transitions from bistable to monostable at \(t^*\), finite-rate driving induces dynamical lag: the system delays switching to the global minimum by \(\hat{t}_\text{del}\), defined by \(m(t^* + \hat{t}_\text{del}) = m^*\). To describe this behavior, we expand the deterministic dynamical equation Eq.~\eqref{SMeq:dmdt} around the critical point $(m^*, H^*)$, leading to the approximate evolution equation~\cite{wuErgodicityBreakingScaling2025}
\begin{equation}
\label{eq:CW-evolution-deterministic-expansion}
\frac{d\hat{m}}{d\hat{t}} = \frac{2\beta J \sqrt{\beta J - 1}}{\tau_0} \hat{m}^2 + \frac{2}{\tau_0} \sqrt{\frac{\beta}{J}} v_H \hat{t},
\end{equation}
where $\hat{m} = m - m^*$ and $\hat{t} = t - t^*$ are the shifted variables, and the delay time follows as~\cite{wuErgodicityBreakingScaling2025}
\begin{equation}
\label{SMeq:tdel}
    \hat{t}_{\text{del}}=-\mathrm{A'_1}\left[4 \sqrt{\beta J(\beta J-1)} \frac{\beta}{\tau_{0}^{2}} v_H\right]^{-1 / 3}
\end{equation}
with the first zero point of the Airy prime function $\mathrm{A'_1}$ (see Appendix~\ref{app:derivation} for details). 
\begin{figure}[htb!]
    \centering
    \includegraphics[width=1\columnwidth]{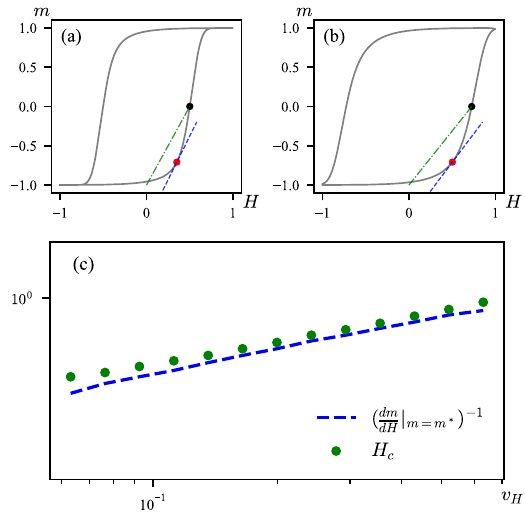}
    
    \caption{(a) and (b)  present numerical observations of two slopes: one at the point 
$(m^*, H^*+v_H \hat{t}_{\text{del}})$, and another linking the coercivity to (0, -1). (c) elaborates on the verification process. Here, we also choose $N=30$,$H_m=1$.}
    \label{SMfig:CW-slope}
\end{figure}

\begin{figure}[htbp]
    \centering
    \includegraphics[width=1\linewidth]{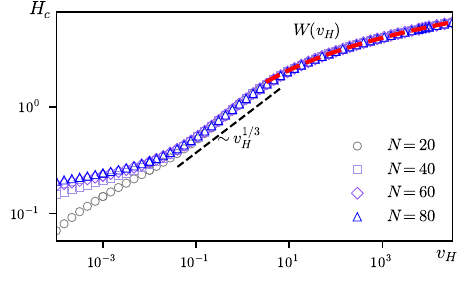}
    \caption{Coercivity of the Curie-Weiss model as a function of driving rate. Detailed scalings in the relatively fast-driving regimes after the plateau, where $W(v_H)$ denotes the Lambert W function (real-valued branch). In this plot, the driving amplitude is unconstrained to ensure full saturation of the order parameter.}
    \label{fig:CW-combined-figure}
\end{figure}

To connect this analytic prediction with our numerical data, we now turn to direct observations of the loop slope at the delayed switching point. Numerical observation suggests that the slope at the point $(m^*, H^*+v_H \hat{t}_{\text{del}})$ corresponding to the delay time $\hat{t}_{\text{del}} $ is approximately equal to the slope connecting the coercivity and the point $(0, -1)$, as shown in Figs.~\ref{SMfig:CW-slope}(a) and (b),  namely,
\begin{equation}
    \frac{dm}{dH}|_{H=H^*+v_H\hat{t}_{\mathrm{del}}}=\frac{\frac{d\hat{m}}{dt}}{v_H}|_{\hat{t}=\hat{t}_{\mathrm{del}},\hat{m}=0}=\frac{1}{H_c}.
\end{equation}
Further numerical verification supports this result, as illustrated in Fig.~\ref{SMfig:CW-slope}(c). By substituting the Airy-function solution Eq.~(\ref{SMeq:tdel}) into Eq.~\eqref{eq:CW-evolution-deterministic-expansion}, the local slope at $\hat m=0$ is obtained as
\begin{equation}
    \frac{d \hat{m}}{d\hat{t}}|_{\hat{t}=\hat{t}_{\text{del}},\hat{m}=0}=-\frac{2\mathrm{A'_1}}{\tau_0}\sqrt{\frac{\beta}{J}}\left[4 \sqrt{\beta J(\beta J-1)} \frac{\beta}{\tau_{0}^{2}} \right]^{-1 / 3}v_H^{2 /3}
\end{equation}
Rearranging the above equations results in the scaling relation
\begin{equation}
    H_c=-\frac{\tau_0}{2\mathrm{A'_1}\sqrt{\beta J}}\left[4 \sqrt{\beta J(\beta J-1)} \frac{\beta}{\tau_{0}^{2}} \right]^{1 / 3}v_H^{1 / 3},
\end{equation}
which emerges between the FOPT plateau and the fast-driving regime. This $H_c \sim v_H^{1/3}$ scaling is marked with the dashed line in Fig.~\ref{fig:CW-combined-figure}, which is universal for different spin numbers. Interestingly, in the faster driving regime, we notice that $H_c$ deviates from the $1/3$-scaling and still exhibit a universal behavior independent of $N$. Next, we analyze the behavior of $H_c$ in this rapidly driven regime. 

\subsubsection{$H_c \sim W(v_H)$}
In the fast-driving regime, the system's dynamics is dominated by the passive relaxation to mono-stable state under strong field, considering that the field increases so fast that the system seems to be frozen before the free energy landscape becomes mono-stable. In the strong-field limit where $|H| \gg |Jm|$, the interaction term $Jm$ becomes negligible compared to $H$, and the deterministic equation~\eqref{SMeq:dmdt} simplifies to
\begin{equation}
    \frac{dm}{dt} \approx \frac{2}{\tau_0} \left[ \sinh(\beta H) - m \cosh(\beta H) \right].
\end{equation}
This linear differential equation describes exponential relaxation toward the equilibrium value
\begin{equation}
    m_{\mathrm{eq}} \approx \tanh(\beta H) \approx 1 - 2 e^{-2 \beta H} \quad \text{for } \beta H \gg 1.
\end{equation}
To estimate the characteristic relaxation time toward equilibrium, we linearize the dynamical equation around $m_{\mathrm{eq}}$:
\begin{equation}
    \frac{dm}{dt} \approx - \frac{2}{\tau_0} \cosh(\beta H)\, (m - m_{\mathrm{eq}}).
\end{equation}
This leads to the relaxation time
\begin{equation}
    \tau_{\mathrm{relax}} \approx \left( \frac{2}{\tau_0} \cosh(\beta H) \right)^{-1} \approx \frac{\tau_0}{2} e^{-\beta H}, \quad \text{for } \beta H \gg 1.
\end{equation}

Unlike comparing the timescales of relaxation and the movement of equilibrium state in the stochastic $\phi^4$ model for analyzing the fast-driving regime, in the Curie-Weiss model, the dominant timescale is only limited by $\tau_{\mathrm{relax}}$ since the equilibrium state infinitely approaches a fixed saturation value in this model. Let $\tau_c$ denote the characteristic time at which the system reaches $m = 0$ during the passive relaxation process to the equilibrium state. 
Considering the relation between the relaxation time and the field strength,
we estimate this time by
\begin{equation}
    \tau_c \sim e^{-\beta H_c},
\end{equation}
Since the external field changes linearly with time, $\tau_c$ can also be estimated by
\begin{equation}
    \tau_c \sim \frac{H_c}{v_H}.
\end{equation}
Combining the above two expressions gives the scaling relation:
\begin{equation}
    H_c e^{\beta H_c} \sim v_H,
\end{equation}
with the Lambert W function being the inverse function of $x e^x$. This scaling behavior is demonstrated by the thick dashed line in Fig.~\ref{fig:CW-combined-figure}. 
The different scaling relations in the fast-driving regime between the $\phi^4$ and Curie-Weiss models due to fundamental differences in their full dynamical equations, highlight the sensitivity of far-from-equilibrium dynamics to model-specific details, even when the underlying physical picture remains similar. Consequently, despite sharing similar physical mechanisms, different microscopic models may reveal more varied scaling relations in this regime ~\cite{raoMagneticHysteresisTwo1990,shuklaHysteresisIsingModel2018}. Furthermore, while the Curie-Weiss model exhibits a sustained $1/3$ scaling that bridges the FOPT characteristics and passive relaxation dynamics at high driving rates, this regime is largely absent in the stochastic $\phi^4$ model, though briefly passed through, which is again attributable to its specific system dynamics.

\subsection{Finite-size scaling of the FOPT plateau}
\label{sec:finite_size_scaling_CW}
In this subsection, we compare the roles of $\sigma$ and $N$ in the Fokker-Planck equations of the stochastic $\phi^4$ and Curie-Weiss models to bridge the gap between stochastic fluctuations and finite-size effects.

We first derive the Fokker-Planck equation corresponding to the master equation~\eqref{SMeq:dPdtCW} for $N\gg1$. Substituting the transition rates~\eqref{SMeq:CW-transition-rate} into Eq.~\eqref{SMeq:dPdtCW}, one obtains
\begin{equation}
\begin{split}
\frac{\partial P(m,t)}{\partial t} &= -\frac{N}{2\tau_0}\left[1-m\right]e^{\beta\left[J\left(m+\frac{1}{N}\right)+H\right]}P(m,t) \\
&\quad - \frac{N}{2\tau_0}\left[1+m\right]e^{-\beta\left[J\left(m-\frac{1}{N}\right)+H\right]}P(m,t) \\
&\quad + \frac{N}{2\tau_0}\biggl[1 - \left(m - \frac{2}{N}\right)\biggr]e^{\beta\left[J\left(m-\frac{1}{N}\right)+H\right]} \\
&\quad \times P\left(m - \frac{2}{N},t\right) \\
&\quad + \frac{N}{2\tau_0}\biggl[1 + \left(m + \frac{2}{N}\right)\biggr]e^{-\beta\left[J\left(m+\frac{1}{N}\right)+H\right]} \\
&\quad \times P\left(m + \frac{2}{N},t\right).
\end{split}
\end{equation}
Using the shift operator~\cite{paulRelaxationMetastableStates1989}, the above equation becomes
\begin{equation}
    \begin{aligned}
        &\frac{\partial P(m,t)}{\partial t}   =\\
        &\frac{N}{2\tau_0}\left(e^{-\frac{2}{N}\frac{\partial}{\partial m}}-1\right)(1-m)e^{\beta\left[Jm+J/N+H\right]}P(m,t) \\
         + &\frac{N}{2\tau_0}\left(e^{\frac{2}{N}\frac{\partial}{\partial m}}-1\right)(1+m)e^{-\beta\left[Jm-J/N+H\right]}P(m,t).
    \end{aligned}
\end{equation}
Considering $N\gg1$, and retaining the $1/N$ term in the above equation, one obtains 
\begin{equation}
\label{eq:CW-FP-eq}
\begin{aligned}
    \frac{\partial P(m,t)}{\partial t} = &- \frac{2+2\beta J/N}{\tau_0} \frac{\partial}{\partial m}\left[T(m,H)P(m,t)\right] \\
    &+ \frac{1}{N}\frac{2}{\tau_0}\frac{\partial^2}{\partial m^2}\left[S(m,H)P(m,t)\right],
\end{aligned}
\end{equation}
where $ T(m,H) = \sinh \beta(Jm+H) - m\cosh \beta (Jm+H)$ and $S(m,H) = \cosh \beta (J m + H) - m \sinh \beta (Jm + H)$. Equation~\eqref{eq:CW-FP-eq} is associated with the deterministic equation~\eqref{SMeq:dmdt} in the $N\to \infty$ limit. Expanding Eq.~\eqref{eq:CW-FP-eq} around the spinodal point $(m^*,H^*)$ given in Eq.~\eqref{SMeq:CW-spinodal-point} and keeping the leading order of each term, we obtain
\begin{equation}
\label{eq:CW-FP-eq-expanding}
\begin{aligned}
    \frac{\partial P(\hat{m},t)}{\partial t}\approx & - \frac{2}{\tau_0}\frac{\partial}{\partial\hat{m}}\left[\beta J\sqrt{\beta J-1}\hat{m}^2+\sqrt{\frac{\beta}{J}}\hat{H}\right]P(\hat{m},t)\\
    &+ \frac{1}{N}\frac{2}{\tau_0}\frac{1}{\sqrt{\beta J}}\frac{\partial ^2}{\partial\hat{m}^2}P(\hat{m},t),
\end{aligned}
\end{equation}
where $\hat{m}\equiv m-m^*$ and $\hat{H}\equiv H-H^*$.

\begin{figure}[htbp]
    \centering
    \includegraphics[width=1\linewidth]{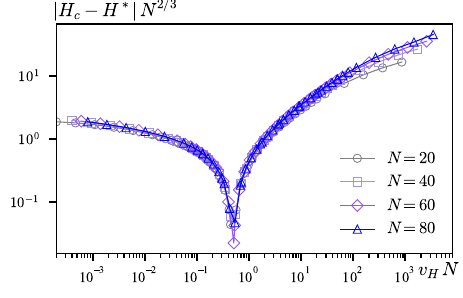}
    \caption{Finite-size scaling collapse of coercivity plateau with a bounded driving amplitude $H_m=1$.}
    \label{CW-scaling-collapse}
\end{figure}

To find the corresponding relation between the system size $N$ in the Curie-Weiss model and the noise strength $\sigma$ in the stochastic $\phi^4$ model, we write down the Fokker-Planck equation for the $\phi^4$ model near its spinodal point,
\begin{equation}
    \frac{\partial P(\varphi,t)}{\partial t} = -\lambda\frac{\partial}{\partial\varphi}\left[\left(-a_3\varphi^2+h\right)P(\varphi,t)\right]+\lambda\sigma^2\frac{\partial^2}{\partial\varphi^2}P(\varphi,t),
\end{equation}
with $\varphi$, $h$ and $a_3$ defined in Eq.~\eqref{eq:phi4-free-energy-phi3}. Comparing the coefficient of the second diffusion part in the above equation and that in Eq.~\eqref{eq:CW-FP-eq-expanding}, one obtains $\sigma^2\sim N^{-1}$. It is worth noting that this relation is independent of the specific transition rates used in the Curie-Weiss model~\cite{moriAsymptoticFormsScaling2010,fioriSpecificHeatDriven2025} and is also observed in different systems~\cite{chenMeasurementNoiseMaximumSignature2007,leffFluctuationsParticleNumber2015}. Therefore, it follows from Eq.~\eqref{eq:phi4-finite-size-scaling-plateau} that
\begin{equation}
    H^* - H_P \propto N^{-2/3} \quad \text{at} \quad v_{P} \sim 1/N.
\end{equation}
According to the above equation, one can rescale the axes in the $H_c-v_H$ plot. As shown in Fig.~\ref{CW-scaling-collapse}, the coercivity curves for different system sizes collapse onto a single universal curve, confirming the validity of the finite-size scaling relation. 

It is indicated from the figure that the universal curve diverges at a critical driving rate that yields $H_c = H^*$. Here $H^*$ is the spinodal field of the Curie-Weiss model. In the infinite-size limit, this critical driving rate approaches zero, rendering the region to its left inaccessible. This demonstrates that the coercivity plateau, which does not vanish in the slow-driving regime, results from taking the thermodynamic limit ($N\rightarrow \infty$) first, namely, $ \lim_{v_H \to 0}\lim_{N\rightarrow \infty}H_c = H^*$. In contrast, for finite-size systems, coercivity values span both sides of this critical point, with the gently varying region at lower driving rates reflecting the presence of a coercivity plateau.

\section{Conclusion}
\label{sec:conclusion} 
In this work, we have developed a unified theoretical framework to characterize coercivity in dynamic hysteresis within the stochastic $\phi^4$ model under periodic driving, spanning the entire spectrum from quasi-static to fast-driving regimes and from finite to infinite system sizes. 

Building on the coercivity panorama proposed in our companion paper~\cite{CP} for the stochastic $\phi^4$ model, we provided detailed analytical derivations and extensive numerical simulations to show the specific morphology of the hysteresis loop  under different noise strengths $\sigma$ and driving rates $v_H$ (see Sec.~\ref{sec:stochastic_phi4}). 

Moving forward, we exhibit the relevant finite-time and finite-size scaling behaviors through precise analytical calculations. Focusing on the coercivity $H_c$ as a succinct measure across noise strengths $\sigma$, we found that, as the driving rate $v_H$ increases under periodic driving, the scaling exponents of coercivity first decrease then increase, verifying the existence of the coercivity plateau. We then calculated two fundamental scaling behaviors: $H_c\sim v_H$ in the near-equilibrium regime and $H_c\sim v_H^{1/2}$ in fast-driving regime in detail. To further characterize the FOPT plateau behavior, we analyze its finite-size scaling via renormalization-group theory and the post-plateau finite-time scaling (see Sec.~\ref{sec:coercivity_stochastic_phi4}). Both of these scaling behaviors are numerically demonstrated via the stochastic $\phi^4$ model in the companion paper~\cite{CP}.

Then we utilized the established coercivity panorama to study the magnetic hysteresis of the
Curie-Weiss model. By solving the governing master equations, we first characterize the specific morphology of the hysteresis loop  under different parameter settings. This enables us to reveal how the loops evolve with varying external conditions including spin number $N$, driving amplitude $H_m$, and driving rate $v_H$. Ultimately, we further analyzed the role of finite system size by constructing the full coercivity panorama for various $N$ (see Sec.~\ref{sec:CW_Dynamics}). In this way, we were able to identify how core characteristics of the coercivity panorama, such as the width of the FOPT plateau and the boundaries between scaling regimes, change with system size.

We also paid attention to the universality of coercivity's finite-time behavior in different driving regimes. Below the FOPT plateau, the linear scaling $H_c\sim v_H$ is universal across both the stochastic $\phi^4$ and Curie-Weiss models. Analogous to the scaling $H_c-H_P\sim(v_H-v_P)^{2/3}$, the Curie–Weiss model exhibits  $H_c-H^*\sim v_H^{2/3}$ in the post-plateau slow-driving regimes. In contrast, despite sharing analogous physical mechanisms, we found that the scaling behavior in the fast-driving regime exhibits clear model-specific differences, with the Curie-Weiss model following the Lambert W-function $W(v_H)$. Moreover, between the FOPT plateau and the fast-driving regime, an intermediate scaling $H_c\sim v_H^{1/3}$ emerges in the Curie-Weiss model, which is briefly traversed without being maintained over an observable timescale in the stochastic $\phi^4$ models (see Sec.~\ref{sec:finite_time_scalings_CW}). These discrepancies highlight the sensitivity of far-from-equilibrium dynamics to model-specific details, even when the underlying physical picture remains similar. Consequently, different microscopic models may reveal more varied scaling relations in this regime.

To bridge the gap between stochastic fluctuations and finite-size effects, we derived the relationship between $\sigma$ and $N$, $\sigma^2\sim N^{-1}$ by comparing the roles of
$\sigma$ and $N$ in the Fokker-Planck equations of the stochastic $\phi^4$ and Curie-Weiss models. We then observed that coercivity for different system sizes collapses onto a single universal curve, thereby confirming the validity of the finite-size scaling relations (see Sec.~\ref{sec:finite_size_scaling_CW}).

The findings of this paper also help clarify the ambiguous interface between experimental observations and theoretical predictions (see the end matter of the companion paper~\cite{CP}). This also inspires further investigations into bridging phenomenological models and fundamental statistical physics frameworks for dynamic hysteresis with specific remarks available in the conclusion of the companion paper~\cite{CP}.


\acknowledgments{We thank Sai Li for the helpful discussions in the early stage of this work. This work is supported by the National Natural Science Foundation of China under grant No. 12305037 and the Fundamental Research Funds for the Central Universities under grant No. 2023NTST017.}

\appendix
\section{Derivation of the approximate equation of motion and the nondimentionalization}
\label{app:derivation}
To isolate the universal structure of the dynamic equation Eq.\eqref{eq:CW-evolution-deterministic-expansion} in the maintext, we introduce dimensionless variables $u = \alpha\hat{m}, s =\gamma\hat{t},$ with the two scale factors $\alpha, \gamma > 0$ chosen to eliminate all prefactors in the resulting equation. In this case, one has
\begin{equation}
\frac{d u}{ds} =\frac{2\beta J\sqrt{\beta J - 1}}{\tau_0\alpha\,\gamma} u^2 + \frac{2\alpha}{\tau_0\gamma^2}\sqrt{\frac{\beta}{J}}v_Hs.
\end{equation}
To rescale the coefficients to unity, we have
\begin{equation}
     \frac{2\beta J\sqrt{\beta J - 1}}{\tau_0\alpha\,\gamma} =1,\quad\frac{2\alpha v_H}{\tau_0\gamma^2}\sqrt{\frac{\beta}{J}}=1,
\end{equation}
 which leads to the following expressions,
\begin{equation}
\begin{aligned}
\alpha &=\left( \frac{2}{\tau_0} \beta J \sqrt{\beta J - 1} \right)^{2/3} \left( \frac{2}{\tau_0} \sqrt{\frac{\beta}{J}} v_H \right)^{-1/3}, \\
\gamma &=\left( \frac{2}{\tau_0} \beta J \sqrt{\beta J - 1} \right)^{1/3} \left( \frac{2}{\tau_0} \sqrt{\frac{\beta}{J}} v_H \right)^{1/3}.
\end{aligned}    
\end{equation}
With these rescaled coefficients, the original equation simplifies to the standard Riccati form
\begin{equation}
\label{SMeq:duds}
\frac{d u}{ds} =u^2 +s,
\end{equation}
Under the asymptotic boundary condition as $s \to -\infty$, $u \to \sqrt{-s}$, the solution to the above equation is the Airy function~\cite{jungScalingLawDynamical1990}
\begin{equation}
u(s) = \frac{\mathrm{Ai}'(-s)}{\mathrm{Ai}(-s)},
\end{equation}
Thus, the delay time is obtained as Eq.\eqref{SMeq:tdel} in the maintext.

\bibliography{refs_for_maintext}

\end{document}